\documentclass{article}

\usepackage{amsmath}
\usepackage{amssymb}
\usepackage{cite}
\usepackage{url}
\usepackage{graphicx}
\usepackage{authblk}

\newcommand{\N}{\mathbb{N}}
\newcommand{\R}{\mathbb{R}}
\newcommand{\T}{\mathbb{T}}

\DeclareMathOperator{\rate}{rate}
\DeclareMathOperator{\arity}{arity}
\DeclareMathOperator{\supp}{supp}

\title{Robustness and Games Against Nature in Molecular Programming}

\author[1]{Jack H. Lutz}
\author[2]{Neil Lutz}
\author[1]{Robyn R. Lutz}
\author[1]{Matthew R. Riley}
\affil[1]{Department of Computer Science, Iowa State University}
\affil[2]{Department of Computer and Information Science, University of Pennsylvania}

\begin{document}

\maketitle

\begin{abstract}
	Matter, especially DNA, is now programmed to carry out useful processes at the nanoscale. As these programs and processes become more complex and their envisioned safety-critical applications approach deployment, it is essential to develop methods for engineering trustworthiness into molecular programs. Some of this can be achieved by adapting existing software engineering methods, but molecular programming also presents new challenges that will require new methods. This paper presents a method for dealing with one such challenge, namely, the difficulty of ascertaining how robust a molecular program is to perturbations of the relative ``clock speeds'' of its various reactions. The method proposed here is game-theoretic. The robustness of a molecular program is quantified in terms of its ability to win (achieve its original objective) in games against other molecular programs that manipulate its relative clock speeds. This game-theoretic approach is general enough to quantify the \emph{security} of a molecular program against malicious manipulations of its relative clock speeds. However, this preliminary report focuses on \emph{games against nature}, games in which the molecular program's opponent perturbs clock speeds randomly (indifferently) according to the probabilities inherent in chemical kinetics. 
\end{abstract}

\section{Introduction}
Molecular programming is, at its simplest, computation with DNA. A programmed molecular system is a nanosystem that will execute the algorithmic behavior encoded into it. Examples of programmed DNA systems include neural net simulation, probabilistic switching circuits, nano-robotic walkers, and oscillators~\cite{jQiWiBr11, jWiBrQi18, cDKTT13, jSPSWS17}. That is, we are programming matter itself when we create a programmed molecular system. 

Many of the intended uses of molecular programming are safety-critical, such as biosensors to detect pollutants, diagnostic devices to identify diseases, and nano-robotic walkers to perform targeted drug delivery~\cite{jDoBaCh12}.
Software engineering techniques including goal-oriented requirements modeling~\cite{oLams09}, risk analyses~\cite{oKnig12}, and probabilistic model checking~\cite{jKwiTha14} have recently been extended to the emerging field of molecular programming, in order to aid the development of safe programmed molecular systems~\cite{cLLLKHM12, jKwiTha14, oEKLLLM17}. 

Assuring the robustness of a molecular program needs to occur before such a system is deployed. However, what robustness means for such a system is not well-defined. This, in turn, hinders efforts to determine how robust a particular system is. 

The problem addressed by this paper is the difficulty of answering the question, ``how robust is this molecular program?'' The contribution of the paper is to propose a game-theoretic method by which we can quantitatively evaluate how robust a molecular program is to an opponent's perturbing the relative clock speeds of its constituent processes (reactions). This sort of robustness is especially important, because the ``rate constants'' that govern the rates of chemical reactions are notoriously approximate and non-constant in actual laboratory experiments. We formulate this as a game against nature~\cite{oMiln51,jPapa85}, in which nature manipulates clock speeds at random, disinterested in the outcome. Although this approach is general enough to evaluate robustness in the face of a game against a malicious opponent, we focus here on the random perturbations inflicted by an indifferent nature. We illustrate our method on an important consensus algorithm, approximate majority~\cite{cCHKM17}. We thus develop a games-against-nature formalism of robustness and then evaluate it on a small molecular program that computes the approximate majority. 

\section{Molecular Programs}

Molecular programs are typically specified as chemical reaction networks (CRNs)~\cite{oAndKur15, oCSWB09}, which are roughly equivalent to stochastic Petri nets~\cite{oDavAll10}. These CRNs can then be automatically compiled into DNA strand displacement systems that can be implemented in laboratory experiments~\cite{jSoSeWi10, jCDSPCS13,cBSJDTW17}. 
 
\subsection*{Syntax}
We now review the definition of CRNs. We fix a countably infinite set $\mathbf{S}$ whose elements are called \emph{species}. We informally regard each species as an abstract type of molecule.

A \emph{reaction} over a finite set $S\subseteq\mathbf{S}$ is an ordered triple
\[\rho = (\mathbf{r},\mathbf{p},k)\in\mathbb{N}^S\times\mathbb{N}^S\times(0,\infty),\]
where $\mathbf{r}\neq\mathbf{p}$ and $\mathbb{N}^S$ is the set of functions from $S$ into $\N$. Given such a reaction $\rho$, we call $\mathbf{r}(\rho)=\mathbf{r}$ the \emph{reactant vector} of $\rho$, $\mathbf{p}(\rho)=\mathbf{p}$ the \emph{product vector} of $\rho$, and $k(\rho)=k$ the \emph{rate constant} of $\rho$. (Since $S$ is finite, it is natural to regard elements of $\N^S$ as vectors.) The species in the \emph{support} $\supp(\mathbf{r})=\{X\in S\mid\mathbf{r}(X)>0\}$ are the \emph{reactants} of $\rho$, and the species in $\supp(\mathbf{p})$ are the \emph{products} of $\rho$.

We usually write reactions in a more chemistry-like notation. For example, if $S=\{X,Y,C\}$, then we write
\[X+C\overset{k}\to 2Y+C\]
for the reaction $(\mathbf{r},\mathbf{p},k)$, where
$\mathbf{r}(X)=1$, $\mathbf{r}(Y)=0$, $\mathbf{r}(C)=1$, $\mathbf{p}(X)=0$, $\mathbf{p}(Y)=2$, and $\mathbf{p}(C)=1$.
A species $C$ satisfying $\mathbf{r}(\rho)(C)=\mathbf{p}(\rho)(C)>0$, as in this example, is called a \emph{catalyst} of the reaction $\rho$. The \emph{net effect} of a reaction $\rho$ is the vector $\Delta\rho=\mathbf{p}(\rho)-\mathbf{r}(\rho)\in\mathbb{Z}^S$.
The \emph{arity} of a reaction $\rho$ is 
\[\arity(\rho)=\sum_{Y\in S}\mathbf{r}(\rho)(Y),\]
i.e., its total number of reactants. A \emph{chemical reaction network (CRN)} is an ordered pair $N=(S,R)$, where $S\subseteq\mathbf{S}$ is finite and $R$ is a finite set of reactions over $S$.

\subsection*{Semantics}
In this paper we assign each CRN $N=(S,R)$ the operational meaning given by the \emph{stochastic mass action semantics} (also called the \emph{stochastic mass action kinetics}) introduced by Gillespie~\cite{jGill77}. In this semantics a \emph{state} of $N$ is a vector $\mathbf{x}\in\mathbb{N}^S$. For each $Y\in S$, the component $\mathbf{x}(Y)$ of $\mathbf{x}$ is the \emph{count} of species $Y$ in the state $\mathbf{x}$. A reaction $\rho$ is \emph{applicable} to state $\mathbf{x}$ if $\mathbf{r}(\rho)\leq \mathbf{x}$, i.e., all the reactants of $\rho$ are present in $\mathbf{x}$. If $\rho$ is applicable to $\mathbf{x}$, then the \emph{result} of applying $\rho$ to $\mathbf{x}$ is the state $\rho(\mathbf{x})=\mathbf{x}+\Delta\rho\in\mathbb{N}^S$.

The \emph{(stochastic mass action) rate} of a reaction $\rho$ in a state $\mathbf{x}\in\N^S$ and volume $V>0$ of solution, which we denote by $\rate_{\mathbf{x}}(\rho)$, was defined and justified by Gillespie~\cite{jGill77}. Here we give a single example. Let $\rho$ be a reaction \[3Y+Z\to\textrm{RHS}.\]
 (The right-hand side RHS does not affect the rate of a reaction.) For brevity, write $y=\mathbf{x}(Y)$ and $z=\mathbf{x}(Z)$. If $\rho$ is applicable to $\mathbf{x}$ (i.e., if $y\geq 3$ and $z\geq 1$), then
 \begin{align*}
 \rate_{\mathbf{x}}(\rho)&=k\cdot V^{1-\arity(\rho)}\cdot y\cdot (y-1)\cdot (y-2)\cdot z\\
 &=ky(y-1)(y-2)z/V^3.
 \end{align*}
Under stochastic mass-action semantics, a CRN $N=(S,R)$ functions as a continuous-time Markov chain~\cite{oRoza69} with state space $\N^S$ and, for each $\mathbf{x},\mathbf{y}\in\N^S$, transition rate
\[\rate(\mathbf{x},\mathbf{y})=\sum_{\rho(\mathbf{x})=\mathbf{y}}\rate_{\mathbf{x}}(\rho).\]
The CRN $N$ is initialized to some state or distribution over states. When it enters a state $\mathbf{x}$, it stays there for a random, real-valued \emph{sojourn time} $t\in(0,\infty]$ before instantaneously executing some reaction $\rho$ and jumping to the state $\rho(\mathbf{x})$. A \emph{trajectory} of $N$ is thus a sequence $\tau=((\mathbf{x}_0,t_0),(\mathbf{x}_1,t_1),\ldots)$
of ordered pairs $(\mathbf{x}_i,t_i)$, where $\mathbf{x}_i$ is a state of $N$ and $t_i$ is the associated sojourn time. The trajectory $\tau$ is finite if it reaches a state to which no reaction is applicable. Otherwise, the trajectory is infinite.

\section{CRN Games}

To quantify the robustness of a CRN, we consider how its performance might be affected by other CRNs that are present in the same solution. Clearly, this evaluation will depend both on how performance is defined and on what kinds of other CRNs are present. We begin by describing a general game-theoretic framework that allows for any scalar quantification of performance (by defining appropriate utility functions) and arbitrary constraints placed on the other CRNs (by restricting the other players' strategy spaces). We then discuss the special case where interactions between CRNs are mediated only by catalysts.

An $n$-player \emph{CRN game} with players $1,2,\ldots,n$ is a pair $\mathcal{G}=(\mathcal{N},\mathbf{u})$ with the following components.

\begin{enumerate}

	\item $\mathcal{N}=\mathbf{N}_1\times\mathbf{N}_2\times\ldots\times\mathbf{N}_n$ is the \emph{strategy profile space}. To play the game, each player $i$ selects a \emph{strategy}: a CRN $N_i=(S_i,R_i)$ from its \emph{strategy space} $\mathbf{N}_i$, which is a set of CRNs. For convenience we require that $(\emptyset,\emptyset)\in\mathbf{N}_i$. The $n$ players' selected strategies collectively define a \emph{strategy profile} $\sigma\in\mathcal{N}$ and a CRN $N_\sigma=(S_\sigma,R_\sigma)$, where $S_\sigma=\bigcup_i S_i$ and $R_\sigma=\bigcup_i R_i$. A \emph{state} of the game comprises counts of all species in $S_\sigma$. As in a CRN, a \emph{trajectory} in this game is a (finite or infinite) sequence of states paired with sojourn times. The space of all trajectories for the game $\mathcal{G}$ is $\T_\mathcal{G}$.

	\item $\mathbf{u}=(u_1,u_2,\ldots,u_n)$ is a profile of \emph{utility functions} $u_i:\T_\mathcal{G}\to\R$,
	where $u_i(\tau)$ is the utility player $i$ gets from trajectory $\tau$. In our example below, this utility function is simply a binary indicator for ``success,'' meaning that the player gets utility 1 from any trajectory that performs a given task correctly and gets utility 0 from all other trajectories. A player representing ``nature'' is totally indifferent to the game's outcome and hence gets utility 0 from all trajectories.
\end{enumerate}

Let $\xi=(\xi_1,\xi_2,\ldots,\xi_n)$ where $\xi_i\in \mathbb{N}^{S_i}$ is the random vector for $N_i$'s initial state, and let $\hat{\xi}_i$ be the embedding of $\xi_i$ in $\mathbb{N}^{S_\sigma}$. Then the initial state of $N_\sigma$ is the random vector $\sum_i\hat{\xi}_i$. Given this initial distribution, the theory of continuous-time Markov chains specifies a probability measure on the set $\T_{N_\sigma}$ of all trajectories of the CRN $N_\sigma$~\cite{oRoza69}, which also immediately yields a probability measure $\mu_{\sigma,\xi}$ on $\T_\mathcal{G}$. This allows us to define the function \[U_i(\sigma,\xi)=\int_{\T_\mathcal{G}}u_i(\tau)\mu_{\sigma,\xi}(\tau),\]
which is the expected value of player $i$'s utility when the strategy profile $\sigma$ is played.

\subsection*{Robustness}
We measure the robustness of a CRN $N_1$ to a profile $(N_2,N_3,\ldots,N_n)$ of other players' CRNs by comparing player 1's expected utility playing $N_1$ against those CRNs to its expected utility playing $N_1$ against trivial CRNs. Formally, for any $\alpha\in[0,1]$, a CRN $N_1\in \mathbf{N}_1$ is \emph{$\alpha$-robust} to $(N_2,N_3,\ldots,N_n)$ in game $\mathcal{G}$ under $\xi$ if
\[U_1((N_1,N_2,\ldots,N_n),\xi)\geq \alpha U_1((N_1,(\emptyset,\emptyset),\ldots,(\emptyset,\emptyset)),\xi'),\]
where $\xi'=(\xi_1,\epsilon,\epsilon,\ldots,\epsilon)$ and $\epsilon$ is the trivial distribution that assigns probability 1 to the empty vector. This means that the participation of other players using these strategies can decrease player 1's expected utility by at most a factor of $\alpha$.

\subsection*{Catalytic Games}
The very general CRN games that we have defined allow essentially unrestricted interactions among the players' CRNs.  For many purposes, including those of this paper, it is more appropriate to restrict these interactions to those mediated by catalysts. 

A \emph{catalytic game} is one in which each player's set of species can be written as $\mathbf{S}_i=\mathbf{A}_i\cup\mathbf{C}_i$ such that
\begin{enumerate}
\item $\mathbf{A}_i$ and $\mathbf{A}_j$ are disjoint for all $j\neq i$, and
\item for all $C\in\mathbf{C}_i$ and $\rho\in\mathbf{R}_i$, we have $\Delta\rho(C)=0$.
\end{enumerate}
In such a game, each player can affect other players only by altering the counts of their catalysts, and hence only by altering the rates of their reactions.

\section{Example}

\subsection*{Approximate Majority}
In this preliminary report, we use a game against nature to investigate the robustness of a simple chemical reaction network that computes approximate majority.

The task in approximate majority is to design a chemical reaction network $N$ with two designated species $X$ and $Y$ and the following objective. Let $x(t)$ and $y(t)$ be the counts of $X$ and $Y$, respectively, at time $t$. First, the total population $x(t)+y(t)$ should be constant as $t$ varies. Moreover, if $x(0)$ and $y(0)$ differ by a non-negligible amount, then whichever is larger should eventually ``take over.'' That is, if $x(0)\gg y(0)$, then we should with high probability have $x(t)=x(0)+y(0)$ (and hence $y(t)=0$) for all sufficiently large $t$. Similarly, if $y(0)\gg x(0)$, then we should with high probability have $y(t)=x(0)+y(0)$ for all sufficiently large $t$. If $x(0)$ is very close to $y(0)$, then we want one of these ``takeovers'' to occur with high probability, but it may be either species that takes over.

We investigate the robustness of the approximate majority CRN
\[R:\begin{array}{l}2X+Y\overset{1}\to 3X\\X+2Y\overset{1}\to3Y\end{array}\]
of Condon, Hajiaghayi, Kirkpatrick and Manuch~\cite{cCHKM17}. In order to do this 
in a catalytic game, we replace $R$ with the catalyzed CRN
\[R':\begin{array}{l}2X+Y+A\overset{1}\to 3X+A\\X+2Y+B\overset{1}\to3Y+B.\end{array}\]
The crucial thing to note here is that, if $x$, $y$, $a$, and $b$ are the counts of $X$, $Y$, $A$, and $B$ at some time, then the rates (``clock speeds'') of the reactions in $R$ at this time are $x(x-1)y$ and $xy(y-1)$, respectively, while the rates of the reactions in $R'$ are $ax(x-1)y$ and $bxy(y-1)$, respectively. If $a$ and $b$ are equal and constant, then $R'$ is merely a uniformly sped-up version of $R$. However, if $a$ and $b$ vary randomly, then the \emph{relative} rates of the reactions in $R'$ also vary randomly (i.e., the ratio of these rates varies randomly).

In order to test the robustness of $R$ to random perturbations of the relative rates of their reactions we thus play the CRN $R'$ against a random ``nature'' that varies the initially equal counts $a$ and $b$ randomly. We model this behavior by the simple CRN
\[N_k:\begin{array}{l}A\overset{k}\to B\\B\overset{k}\to A\end{array},\]
which is ``calibrated'' by the rate constant $k\in(0,\infty)$.

\subsection*{Simulation}
We assessed the robustness of the approximate majority CRN $R$ by comparing the performance of the catalyzed CRN $R'$ in isolation with its performance in the presence of the CRN $N_k$ representing nature randomly perturbing rate constants. We frame this as a game where the utility of the approximate majority player is given by its success frequency. 

We first created models in MATLAB using SimBiology software tools for the catalyzed approximate majority CRN $R'$ and the nature CRN $N_k$ described above. With initial populations $a(0)=b(0)=100$ and combined initial population $x+y=10,000$, we varied the difference $x(0)-y(0)$ from 0 to 1,000 by intervals of 10. With these initial conditions, we ran $R'$ in the presence of $N_k$ with $k=10^9$. We ran 10,000 trials for each set of initial conditions. Each trial converged within $10^{-8}$ time units, meaning that either $x(10^{-8})$ or $y(10^{-8})$ was 0. The design of the CRN $R'$ guarantees that, once one population has taken over, no further reactions can occur. 

\begin{figure}[h]
	\includegraphics[width=\columnwidth]{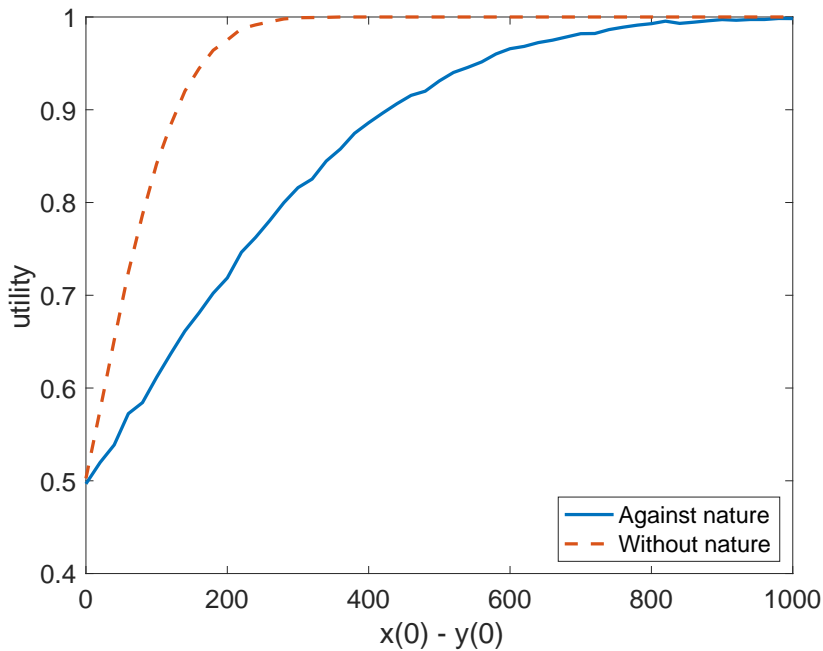}
	\caption{Success frequency of the approximate majority CRN $R'$ with combined initial populations $x(0)+y(0)=10,000$ and initial catalyst populations $a(0)=b(0)=100$, both in isolation and in a game against the nature CRN $N_{10^9}$, for varying values of $x(0)-y(0)$.}
	\label{fig:simresults}
\end{figure}

As Fig.~\ref{fig:simresults} shows, the random perturbations of $a$ and $b$ did reduce the success probability of the approximate majority algorithm for some values of $x(0)-y(0)$. For example, when $x(0)=5,120$ and $y(0)=4,880$, $R'$ was $99\%$ successful in a vacuum but only $76\%$ successful in the presence of nature. However, even with the random perturbations, the success frequency in the presence of nature was always greater than $70\%$ of the success frequency in the absence of nature.
This suggests that the CRN $R'$ is at least $0.7$-robust to $N_{10^9}$ in this game under arbitrary distributions of initial states with $a=b=100$ and $x+y=10,000$.

\section{Conclusion}
Software engineering for molecular programming is a new research direction with open problems that can benefit from the attention of the software engineering research community. Many planned molecular systems will be deployed for use \emph{in vivo} within a few years, and certification for safety-critical scenarios such as biosensors and drug delivery devices will require improved evidence of robustness. Molecular program developers similarly will be called upon to prevent system design vulnerabilities to malicious adversaries. Software engineering has an essential role to play in what scientists are already labeling as the century of life sciences~\cite{jFiHaHe11}.

The preliminary work described in this paper uses a game-theoretic approach to (1) formulate the robustness of a molecular program's CRN model in terms of a game against nature and (2) provide a method to quantitatively evaluate its robustness. The example we present concerns random perturbations of the program's clock speed by nature; however, the approach is general enough to also enable evaluation of security against an adversarial molecular program who maliciously perturbs the relative clock speeds. Future work on this will entail challenging issues involving strategic equilibria~\cite{jNeym17}. Our approach provides a foundation from which to pursue improved development and deployment of verifiably robust programmed molecular systems. More broadly, robustness in the presence of probabilistic behavior also is required for many non-molecular programmed systems~\cite{cSCFR17, cFiGhTa11, jCGKM12, jPaBrUc16}, and the advances described here may enhance our understanding of how to design in and verify robustness for other asynchronous systems operating in stochastic environments.

\section*{Acknowledgments}
This research was supported in part by National Science Foundation grant 1545028. We thank Jim Lathrop for tool assistance.

\bibliographystyle{plain}
\bibliography{master}

\end{document}